# A Computationally Efficient Hybrid Neural Network Architecture for Porous Media: Integrating Convolutional and Graph Neural Networks for Improved Property Predictions

Qingqi Zhao[1], Xiaoxue Han[2], Ruichang Guo[3], and Cheng Chen[4*]

1 Stuttgart Center for Simulation Science (SC SimTech), University of Stuttgart, Stuttgart, Germany.

2 Department of Computer Science, Stevens Institute of Technology, Hoboken, NJ, USA.

3 The Bureau of Economic Geology, University of Texas at Austin, Austin, Texas, USA.

4 Department of Civil, Environmental and Ocean Engineering, Stevens Institute of Technology, Hoboken, NJ, USA.

*Corresponding author; email: cchen6@stevens.edu

**Keywords:** Deep learning; Permeability; Porous media; Pore network modeling; Graph neural network

## Abstract

Porous media is widely distributed in nature, found in environments such as soil, rock formations, and plant tissues, and is crucial in applications like subsurface oil and gas extraction, medical drug delivery, and filtration systems. Understanding the properties of porous media, such as the permeability and formation factor, is crucial for comprehending the physics of fluid flow within them. We present a novel fusion model that significantly enhances memory efficiency compared to traditional convolutional neural networks (CNNs) while maintaining high predictive accuracy. Although the CNNs have been employed to estimate these properties from high-resolution, three-dimensional images of porous media, they often suffer from high memory consumption when processing large-dimensional inputs. Our model integrates a simplified CNN with a graph neural network (GNN), which efficiently consolidates clusters of pixels into graph nodes and edges that represent pores and throats, respectively. This graph-based approach aligns naturally with the porous medium structure, enabling large-scale simulations that are challenging with traditional methods. Furthermore, we use the GNN Grad-CAM technology to provide new interpretability and insights into fluid dynamics in porous media. Our results demonstrate that the accuracy of the fusion model in predicting porous medium properties is superior to that of the standalone CNN, while its total parameter count is nearly two orders of magnitude lower. This innovative approach highlights the transformative potential of hybrid neural network architectures in advancing research on fluid flow in porous media.

## 1. Introduction

Fluid flow in porous media plays a crucial role in various natural and engineered processes, including contaminant transport in groundwater, subsurface nuclear waste disposal, and geological carbon sequestration (R. Guo et al., 2021; McCarthy & Zachara, 1989; Teng et al., 2023). Recent advancements in imaging technologies, such as X-ray micro-computed tomography, have made it possible to visualize the internal structures of porous media in three-dimensional (3D) space. These technologies facilitate the characterization of their physical properties, such as permeability, through numerical simulations (D. Zhang et al., 2000; Zhao et al., 2019).

Traditional approaches for predicting permeability from these digital images involve direct numerical simulations (DNS) based on solving governing equations, such as Darcy's law or the Navier-Stokes equations, to model fluid flow through porous materials. DNS methods, including the finite element method (Karim et al., 2014), finite volume method (Jaganathan et al., 2008), and the lattice Boltzmann (LB) method (R. Guo et al., 2020; Z. Guo et al., 2002; Kang et al., 2007), are known for their high computational cost, particularly when applied to high-resolution 3D images. To make computations more effecient, upscaling techniques, such as pore network modeling (PNM) (Blunt, 2001; Blunt et al., 2013; Joekar-Niasar & Hassanizadeh, 2012), have been developed. PNM simplifies the complex pore structure into a network of interconnected pores and pore throats, allowing efficient yet approximate predictions of permeability. However, these methods often rely on simplifying assumptions, which can impact the accuracy of the permeability prediction.

The recent advancements in neural network models have significantly propelled research in pattern recognition and image segmentation (LeCun et al., 2015; Nguyen et al., 2023). Deep learning technologies, particularly convolutional neural networks (CNNs), offer a promising method for efficiently and accurately estimating porous medium properties (Alqahtani et al., 2021; Hong & Liu, 2020; Kamrava et al., 2020; Liu et al., 2023; Mudunuru et al., 2022; Tembely et al., 2020). Previous studies have demonstrated the potential of CNNs as surrogates for image-based predictions of porous medium properties. Modifications to the model architecture and the incorporation of physical information have been employed to enhance the accuracy of the predictions (Al-Zubaidi et al., 2023; Fu et al., 2023; Kang et al., 2024; Marcato et al., 2022; Tang et al., 2022; Y. D. Wang et al., 2021; Zhai et al., 2024a). For example, J. Wu et al. (2018) utilized a CNN with physical parameters for permeability predictions, achieving better results than regular CNNs. Elmorsy et al. (2022) designed a CNN architecture with an inception module featuring two parallel paths, improving the model's generalizability by using different kernel sizes in the parallel paths to predict unseen rock samples with high accuracy. H. Zhang et al. (2022) developed an autoencoder-based CNN for low-resolution permeability prediction, demonstrating significantly better results than lattice Boltzmann simulations when using low-resolution images as the input. Kamrava et al. (2021) developed a deep CNN for estimating the dispersion coefficient in flow through heterogeneous porous media, showing rapid and accurate predictions of porous medium properties. Other methods like generative adversarial networks (GANs) have also been applied to data-driven research in porous media (Ershadnia et al., 2024; Ferreira et al., 2022; Qian et al., 2024). Umanovskiy (2022) applied a GAN architecture to model two-phase flow in porous media, achieving results comparable to traditional numerical simulations with data output rates two to three orders of magnitude faster. Z. Wang et al. (2022) utilized a conditional GAN for pore-scale modeling of two-phase flow in porous media, effectively predicting fluid saturation and spatial distribution across varying wetting conditions and particle shapes.

In addition to the direct use of data-driven models, physics-informed neural networks (PINNs) have also been applied to porous medium property prediction (Ebadi et al., 2021; Kashefi & Mukerji, 2023; Xu et al., 2023). Elmorsy et al. (2023) developed a CNN model equipped with a physics-informed module that achieves excellent accuracy, even with unseen porous media samples. Almajid & Abu-Al-Saud (2022) utilized a PINN to predict the physics of fluid flow in porous media, showing high accuracy when coupled with observed data.

Additionally, attempts have been made to utilize neural networks for the prediction of flow fields in porous media, which can offer more comprehensive insights than upscaling porous medium properties (Da Wang et al., 2020; Kamrava, Sahimi, et al., 2021; Ko et al., 2023; Santos et al., 2020, 2021; X.-H. Zhou et al., 2022). Kashefi & Mukerji (2023) introduced a physics-informed PointNet (PIPN) to predict Stokes flows in porous media, which improved memory efficiency and reduced computational costs. Yan et al. (2022) introduced a gradient-based deep neural network model that integrated differential operators from governing partial differential equations to simulate multiphase flows in porous media, achieving high accuracy in predicting nonlinear flow dynamics while leveraging physics-informed regularization.

While CNNs can effectively identify geometric features, they may lose certain details during the convolution process with activation functions (Tang et al., 2022). Moreover, CNNs are primarily designed for regular Euclidean data, such as images (two-dimensional (2D) grids) and text (one-dimensional sequences). Adapting CNNs to non-Euclidean domains presents challenges, particularly in defining localized convolutional filters and pooling operators. Additionally, CNNs can be memory intensive when processing 3D images. Prior approaches using CNN for 2D image recognition problems often struggled to scale up to large 3D domains required for a representative elementary volume (REV). This limitation caused many studies to focus on homogeneous samples with smaller REVs, where the global nature of fluid flow could be neglected and the memory bottleneck could be bypassed (Santos et al., 2021). To overcome these challenges, geometric deep learning, particularly deep learning on graphs, has gained significant attention as an emerging research area (Bronstein et al., 2017).

One advancement in addressing these challenges is the development of graph neural networks (GNNs), a specialized neural network designed for graph data structures (Chen et al., 2016; Hamilton et al., 2018; Kipf & Welling, 2017; Vashishth et al., 2020; J. Zhou et al., 2020). GNNs were developed to address the limitations of traditional CNNs in handling the irregular sizes and complex structures of graphs. GNNs are adept at capturing the complex relationships between nodes in a graph, leading to significant advancements in graph analysis research. Their applications include a wide range of tasks, such as graph classification, link prediction, community detection, and graph embedding (Z. Wu et al., 2021). Recently, GNNs have risen in popularity across various domains, from social networks and recommender systems to life sciences and biomedical engineering. In molecular property predictions, for example, the graph architecture is naturally aligned with the input representations of molecules and materials, which can be viewed as chemical graphs of atoms and bonds (Jones et al., 2021). This alignment allows GNNs to access comprehensive atomic-level material representations (Li et al., 2022). By incorporating physical laws and large-scale phenomena, GNNs can develop detailed material representations that are useful for specific tasks, such as predicting material properties (Reiser et al., 2022). As a result, GNNs offer a promising alternative to traditional feature representations commonly used in the natural sciences.

Despite the potential of GNNs, their application in predicting porous medium properties has not been extensively explored in previous research. Alzahrani et al. (2023) presented a hybrid deep learning framework that combines GNNs with CNNs to predict porous medium properties. They treated each sample as a node in the training graph. Jiang & Guo (2024) constructed surrogate models based on graph convolutional networks to approximate the spatial-temporal solutions of the multi-phase flow and transport processes in porous media with unstructured meshes. The results demonstrated that the surrogates accurately predicted the evolutions of pressure and saturation states. Different from prior studies, we directly represent pore structures using graphs extracted from digital rock images through PNM, enabling an efficient approach to capturing intricate relational data between pores. Our study aims to develop and evaluate a novel GNN-CNN fusion model that leverages the strengths of both CNNs and GNNs for enhanced multiscale feature extraction and efficient predictions of porous medium properties. The GNN component consolidates clusters of pixels into nodes and edges that represent pores and pore throats, respectively, which align naturally with the porous medium structure and enhance computational efficiency. Furthermore, we employ the GNN Grad-CAM technology to provide interpretability and novel insights into fluid dynamics within porous media. By integrating these approaches, our model not only achieves improved prediction accuracy but also enables more scalable simulations of complex structures. This work represents a pioneering effort in digital rock physics, showcasing the transformative potential of hybrid neural network architectures in property predictions and modeling of porous media.

## 2. Methods

### 2.1. Data Extraction and Structure

The majority of research on the implementation of deep learning technologies for the prediction of digital rock properties is based on the CNN models (Y. D. Wang et al., 2021). CNNs are well-suited to address issues related to spatial correlations, given their capacity to recognize shift-invariance, local connectivity, and compositionality of image data (Bronstein et al., 2017). As illustrated in **Fig. 1a**, a CNN framework for the analysis of digital rock data typically uses images as the inputs, represented as regular grids in Euclidean space. In this way, CNNs can extract meaningful local features that are shared with the entire data sets for the analysis of porous media images. However, a significant challenge for CNNs is the increased computational and memory demands during training, especially for large 3D images, where the escalating number of pixels significantly increases memory usage, consequently diminishing the model's computational speed. Numerous studies have highlighted the constraints of current computational resources in forecasting rock characteristics via CNNs (Elmorsy et al., 2022; Y. D. Wang et al., 2021).

Our dataset was sourced from the DeePore framework (Rabbani et al., 2020), consisting of 17,700 semi-realistic 3D microstructures of porous geomaterials, each with dimensions of $256^3$ voxels. The original core of this dataset was derived from 60 real microtomography images collected from various geological formations. Detailed information about the sampling locations and characteristics of these geological formations can be found in Appendix A of the DeePore study. To generate a comprehensive and diverse dataset, several data augmentation techniques were employed, such as elastic deformation, image mixing, and cross-correlation-based simulation. To ensure the quality and relevance of the data, rigorous data cleaning and pre-processing steps were carried out. Non-physical or outlier geometries, such as samples without percolating pathways, which can result in zero hydraulic permeability or infinite formation factors, were filtered out. Additionally, normalization was applied to the Euclidean distance transforms to scale the values between -1 and 1, ensuring uniformity across the dataset.

To address the computational limitations of processing full 3D images, DeePore dataset transformed the 3D images into 2D representations. Specifically, three perpendicular mid-planes, in the x-y, y-z, and x-z directions, were extracted from each 3D volume. The resulting 2D images were then formatted into the standard CNN input dimensions, where the three planes served as separate input channels.

Porous media can be considered a kind of graph due to their pore-throat-pore connection relationship. While CNNs have proven to be effective in capturing hidden patterns in Euclidean data, the inherent complexity of graph data has posed significant challenges to the existing machine learning algorithms. In contrast to the CNN model, GNNs employ distinct methodologies for data analysis. In the context of digital rock studies, GNNs facilitate the conversion of digital rock images into graph formats that represent the pore and throat structure as nodes and edges (**Fig. 1b**). This approach contrasts with the CNN approach of examining digital rock images at the pixel level. Our study employed a watershed segmentation algorithm, as implemented in PoreSpy (Gostick et al., 2019), to generate the PNM from the images. The PNM is a widely utilized conceptual framework in the study of porous media, offering a simplified representation of the intricate microscopic architecture (Blunt, 2001). This method uses the Euclidean distance transformation to segment the microstructure, identifying individual pores and throats, which are then represented as graph nodes and edges, respectively. Properties such as pore volume and throat diameter are computed during this process and then assigned to the nodes and edges, respectively. This transformation allows for the scaling up of the input data from the pixel level to the pore level, which has the effect of significantly reducing memory usage and accelerating the training process. It should be noted that graphs may be irregular in structure, with each graph potentially comprising a different number of nodes. Furthermore, the number of neighbors that individual nodes possess may vary within a given graph. Such a flexible data structure endows the GNN with a more adaptable data input dimension, eliminating the need for fixed input data dimensions as seen in CNNs. This representation enables GNNs to concentrate on the interconnectivity and interrelationships within the data, which is of significant importance for comprehending the intricacies of the pore structure.

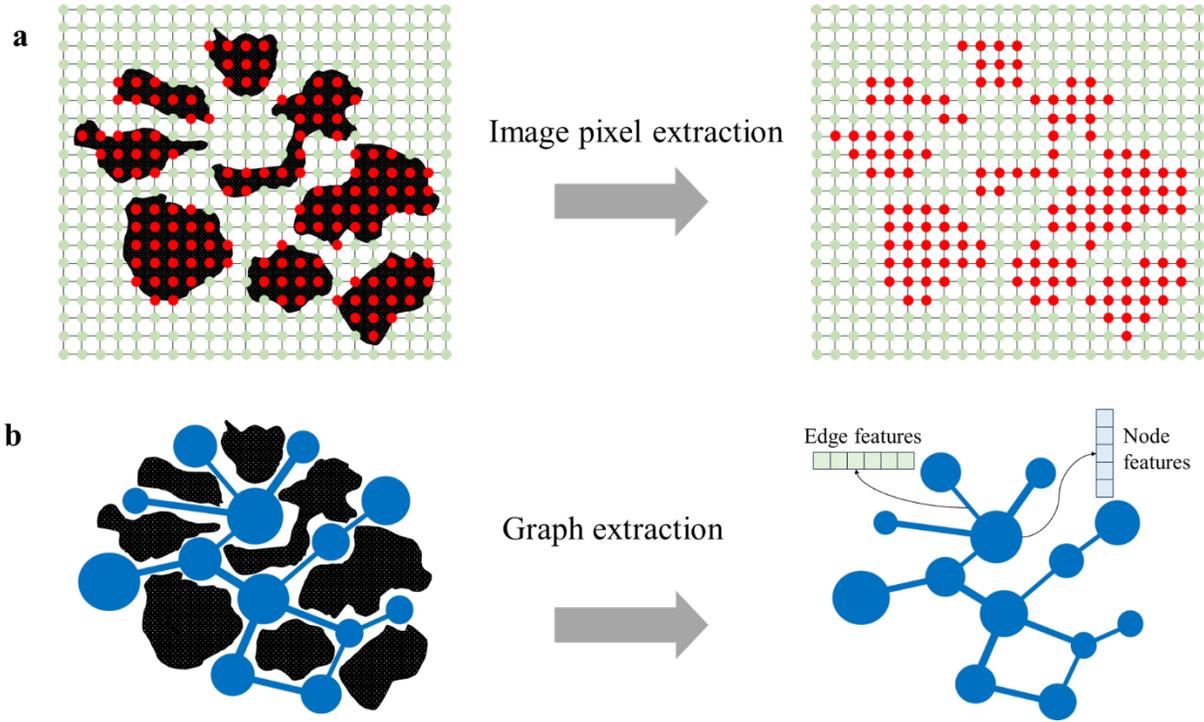

**Fig. 1**. Illustration of data extraction from digital rock images utilizing the CNN and GNN. (a) Pixel extraction from a digital rock image. (b) Graph extraction from a PNM, with the node and edge features sourced from PNM calculations.

The features of each node and edge in the graph are summarized in **Table 1**. Using various diameter measurements, such as inscribed, equivalent, and extended diameters, is a significant advantage of our GNN-based approach. A realistic pore structure cannot be accurately represented by a single type of diameter measurement. By incorporating multiple types of diameters, our GNN approach captures the complex geometry of the porous media more effectively. Another essential feature of our model is the bi-directionality of the edges, ensuring that information is not just passed in a single direction but can flow both ways between nodes. This bi-directional flow facilitates a more robust and comprehensive understanding of the interconnectedness and interactions within the porous media.

**Table 1.** Summary of graph input features.

| Feature type | Properties | Description |
|---|---|---|
| **Node feature** | | |
| | Pore volume | The space or void within the pore, calculated from the pore diameter assuming a spherical pore body |
| | Pore equivalent diameter | A calculated diameter representing pore size based on volume, determined as the diameter of a sphere |
| | Pore inscribed diameter | The diameter of the largest sphere that can fit inside the pore |
| | Pore extended diameter | Accounts for extensions or irregularities in the pore shape |
| | Pore surface area | The internal surface area of pore bodies, calculated by assuming a circular shape and subtracting the area of neighboring throats |
| **Edge feature** | | |
| | Throat inscribed diameter | The largest circle diameter that can fit within the throat |
| | Throat total length | The entire length of the throat |
| | Throat direct length | The shortest path or straight-line distance within the throat |
| | Throat perimeter | The boundary length of the throat, calculated assuming a circular, square, or rectangular cross-section |
| | Throat cross-sectional area | The area available for flow within the throat |
| | Throat equivalent diameter | A diameter based on the throat's cross-sectional area |

The variability of the pore and throat properties is illustrated in **Fig. 2 and 3**, which present the average value distributions for each graph feature across the dataset. These figures provide a representative overview of the data, capturing central tendencies and general patterns for each property. The variability observed in these distributions reflects the diverse geometric characteristics of the porous media samples.

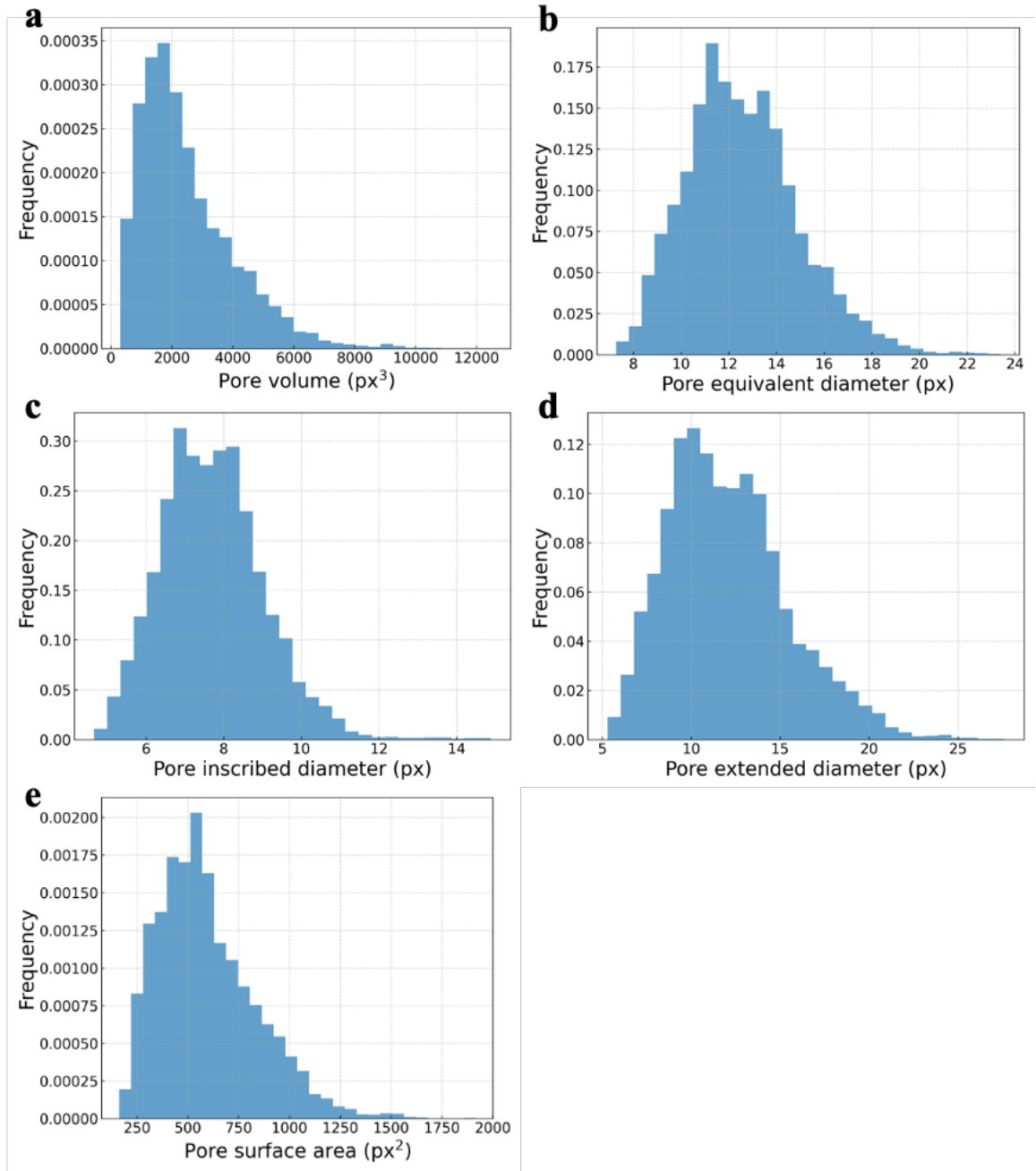

**Fig. 2.** Distributions of pore properties of the dataset: (a) Pore volume, (b) pore equivalent diameter, (c) pore inscribed diameter, (d) pore extended diameter, and (e) pore surface area. The histograms are normalized to represent frequency distributions.

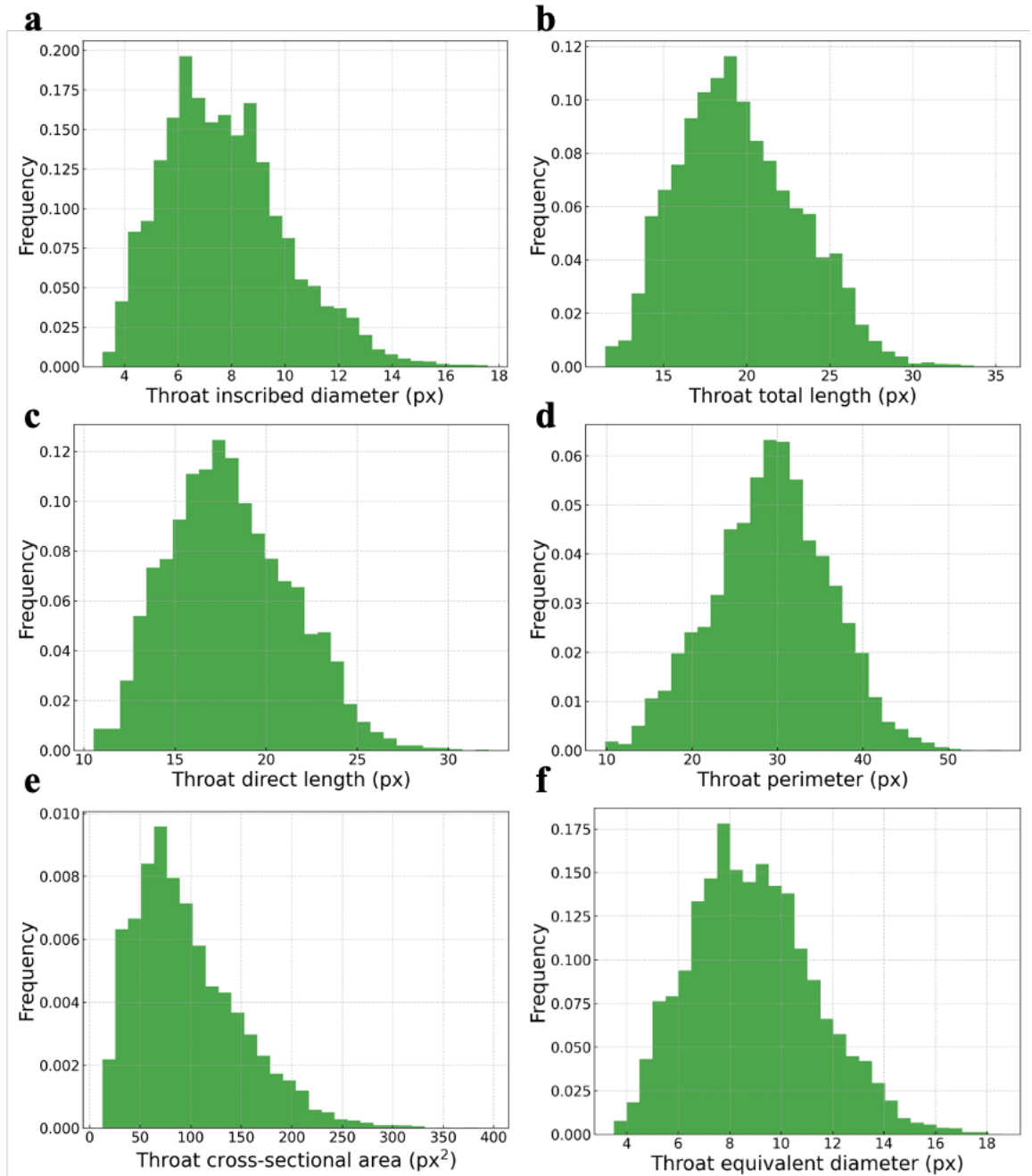

**Fig. 3**. Distributions of throat properties of the dataset: (a) Throat inscribed diameter, (b) throat total length, (c) throat direct length, (d) throat perimeter, (e) throat cross-sectional area, and (f) throat equivalent diameter. The histograms are normalized to represent frequency distributions.

The $256^3$ voxels images from dataset DeePore were utilized as the inputs to CNNs, and their extracted graphs were subsequently employed as the inputs to the GNN. The DeePore dataset features 30 physical properties of each sample calculated using physical simulations on the corresponding PNMs. In our study, we utilized the permeability and formation factor values from the physical property dataset as the ground truth for our deep learning models. The lengths within this dataset are measured in the unit of pixel (px). As a result, permeabilities for all synthetic geometries are reported in the unit of pixel squared ($px^2$). The rescaled permeability value can be determined by multiplying the reported permeability in $px^2$ by the square of the spatial resolution.

## 2.2. Physics of Porous Media

In this study, we utilized two porous medium properties, permeability and formation factor as the ground truth for training the neural network models. Both permeability and formation factor are highly influenced by the structures of the porous media.

Permeability is a property of porous materials that indicates the ability of fluids (gas or liquid) to flow through them. Fluids can more easily flow through a material with high permeability than through one with low permeability. The permeability of a medium is related to porosity, but also to the shapes of the pores in the medium and their level of connectivities. It can be described by Darcy's law:

$$Q = \frac{kA\Delta P}{\mu L} \tag{1}$$

where $Q$ is the volumetric flow rate, $k$ is the permeability of the medium, $A$ is the cross-sectional area perpendicular to the flow direction, $\Delta P$ is the pressure drop across the medium, $\mu$ is the dynamic viscosity of the fluid, $L$ is the length of the medium in the flow direction. To calculate permeability, we rearrange Darcy's law as follows:

$$k = \frac{Q\mu L}{A\Delta P} \tag{2}$$

The formation factor quantifies the impact of pore space on the resistance of the sample. It helps describe the electrical behavior of a porous medium saturated with a conductive fluid. The formation factor is a dimensionless number that represents the ratio of the electrical resistance of the fully saturated porous media to the electrical resistance of the pure fluid. For clean (clay-free) sandstones and similar materials, Archie's law (Adler et al., 1992) provides a relationship between formation factor, porosity, and the properties of the porous medium:

$$F = a \cdot \emptyset^{-m} \tag{3}$$

where $F$ is the formation factor, $a$ is the tortuosity factor, $\emptyset$ is the porosity of the medium, and $m$ is the cementation exponent (a constant that depends on the rock type). Both permeability and formation factor are critical parameters of porous media and are highly related to porous media structure (Rabbani et al., 2020). Therefore, we used these two porous medium properties in this study for the model evaluation.

## 2.3. Deep Learning Models

We implemented and compared two CNN architectures: the 2D CNN and the 3D CNN. The primary distinction between the two CNN models lies in their input data dimensions. For the 3D CNN, we utilized the original images having $256^3$ voxels. In contrast, the 2D CNN model transformed the 3D images into 3 three-channel 2D images. In our study, the 2D CNN, 3D CNN, and hybrid GNN-2D CNN models share a structurally similar hierarchical design, with the primary differences being the input dimensions and the integration of graph-based features. All models are constructed using the PyTorch framework and incorporate alternating convolutional layers and pooling operations, culminating in fully connected layers for the output. The detailed architectures of these models are presented in Tables 2, 3, and 4.

**Table 2.** 3D CNN model architecture.

| Layer type | Input dimension | Kernel size | Activation function | Output dimension |
|---|---|---|---|---|
| Convolutional | 256×256×256×1 | 3×3×3 | ReLU | 256×256×256×3 |
| Max-Pooling | 256×256×256×3 | 2×2×2 | - | 128×128×128×3 |
| Convolutional | 128×128×128×3 | 3×3×3 | ReLU | 128×128×128×6 |
| Max-Pooling | 128×128×128×6 | 2×2×2 | - | 64×64×64×6 |
| Convolutional | 64×64×64×6 | 3×3×3 | ReLU | 64×64×64×12 |
| Max-Pooling | 64×64×64×12 | 2×2×2 | - | 32×32×32×12 |
| Flatten | 32×32×32×12 | - | - | 393,216 |
| Dropout | 393,216 | - |  | 393,216 |
| Fully connected | 393,216 | - | ReLU | 128 nodes |
| Fully connected | 128 nodes | - | ReLU | 64 nodes |
| Output | 64 nodes | - | - | 1 node |

**Table 3.** GNN model architecture.

| Layer type | Input dimensions | Operation details | Output dimensions |
|---|---|---|---|
| GNNLayer (conv1) | node features = 6, edge features = 6 | Linear transformation, message passing, ReLU activation | hidden features1 = 12 |
| GNNLayer (conv2) | hidden features1 = 12, edge features = 6 | Linear transformation, message passing, ReLU activation | hidden features2 = 12 |
| Global Mean Pool | hidden features2 = 12 | Aggregates features across graph nodes | hidden features2 = 12 |
| BatchNorm | hidden features2 = 12 | Batch normalization | hidden features2 = 12 |
| Fully Connected | hidden features2 = 12 | Linear transformation, ReLU activation | predictor hidden features = 128 |

| | Fully Connected | predictor hidden features = 128 | Linear transformation | output = 1 |

**Table 4.** Hybrid GNN-2DCNN model architecture.

| Component | Layer type | Input dimensions | Operation details | Output dimensions |
| --- | --- | --- | --- | --- |
| **GNN Component** | Refer to Table 3 | Refer to Table 3 | Refer to Table 3 | hidden features2 (128) |
| **2D CNN Component** | | | | |
| | Convolutional | 128×128×3 | 3×3 kernel, ReLU activation | 128×128×6 |
| | Max-Pooling | 128×128×6 | 2×2 pooling | 64×64×6 |
| | Convolutional | 64×64×6 | 3×3 kernel, ReLU activation | 64×64×12 |
| | Max-Pooling | 64×64×12 | 2×2 pooling | 32×32×12 |
| | Convolutional | 32×32×12 | 3×3 kernel, ReLU activation | 32×32×24 |
| | Max-Pooling | 32×32×24 | 2×2 pooling | 16×16×24 |
| | Flatten | 16×16×24 | - | 6144 |
| | Dropout | 6144 | - | 6144 |
| | Fully Connected | 6144 | Linear transformation, ReLU activation | 128 |
| | Fully Connected | 128 | Linear transformation, ReLU activation | 64 |
| **Fusion & MLP** | | | | |
| | Concatenate | GNN (128) + CNN (64) | - | 192 |
| | Fully Connected | 192 | Linear transformation, ReLU activation | 64 |
| | Fully Connected | 64 | Linear transformation, ReLU activation | 32 |

| | Output | 32 | Linear transformation | 1 |

Our GNN model is specifically designed to process graphs extracted from the 3D images. In GNNs, each node $v$ in a graph $G = (V, E)$ is represented by a feature vector $\mathbf{h}_v \in \mathbb{R}^F$, where $F$ is the number of node features. Each edge $e \in E$ has an associated feature vector $\mathbf{e}_{uv} \in \mathbb{R}^E$. The relationships between nodes are captured by an adjacency matrix $\mathbf{A} \in \mathbb{R}^{N \times N}$, where $\mathbf{A}_{uv} = 1$ if there is an edge between nodes $u$ and $v$, and $\mathbf{A}_{uv} = 0$ otherwise. To include self-loops and ensure numerical stability, we add the identity matrix $I$ to the adjacency matrix $\mathbf{A}$:

$$\widetilde{\mathbf{A}} = \mathbf{A} + \mathbf{I} \tag{4}$$

We then compute the normalized adjacency matrix:

$$\widehat{\mathbf{A}} = \widetilde{\mathbf{D}}^{-\frac{1}{2}} \widetilde{\mathbf{A}} \widetilde{\mathbf{D}}^{-\frac{1}{2}} \tag{5}$$

where $\widetilde{\mathbf{D}}$ is the degree matrix for $\widetilde{\mathbf{A}}$:

$$\widetilde{\mathbf{D}}_{ii} = \sum_j \widetilde{\mathbf{A}}_{ii} \tag{6}$$

The degree matrix $\widetilde{\mathbf{D}}$ is a diagonal matrix where each diagonal element $\widetilde{\mathbf{D}}_{ii}$ represents the degree of node $i$, i.e., the number of edges connected to node $i$ (including self-loops). This matrix is crucial for normalizing the adjacency matrix, which helps prevent numerical instabilities and ensures the stability of the training process. GNN utilizes a message passing mechanism to aggregate information from neighboring nodes and edges. The message passing process can be mathematically represented as:

$$\mathbf{m}_v^{(l)} = \sum_{u \in \mathbb{N}(v)} \sigma\left(\mathbf{W}_1^{(l)} \mathbf{h}_u^{(l)} + \mathbf{W}_e^{(l)} \mathbf{e}_{uv} + \mathbf{b}_1^{(l)}\right) + \sigma\left(\mathbf{W}_2^{(l)} \mathbf{h}_v^{(l)}\right) \tag{7}$$

$$\mathbf{h}_v^{(l+1)} = \sigma\left(\mathbf{W}_3^{(l)} \mathbf{m}_v^{(l)} + \mathbf{b}_2^{(l)}\right) \tag{8}$$

where $\mathbf{m}_v^{(l)} \in \mathbb{R}^{F'}$ represents the aggregated message for node $v$ at layer $l$ and $\mathbf{h}_v^{(l)} \in \mathbb{R}^F$ represents the updated feature vector of node $v$ at layer $l$. $\mathbf{W}_1^{(l)}, \mathbf{W}_2^{(l)} \in \mathbb{R}^{F \times F'}$ $\mathbf{W}_e^{(l)} \in \mathbb{R}^{E \times F'}$, and $\mathbf{W}_3^{(l)} \in \mathbb{R}^{F \times F'}$ are the layer-specific trainable weight matrices. $\mathbf{b}_1^{(l)}, \mathbf{b}_2^{(l)} \in \mathbb{R}^{F'}$ are the bias vectors. $\sigma$ is a non-linear activation function; in this work, we utilize ReLU.

Based on the message passing mechanism, a multi-layer GNN model is built to capture higher-order relationships between nodes. The input to the first GNN layer is $\mathbf{X} \subseteq \mathbb{R}^{N \times d}$. The output is the intermediate node embeddings $\mathbf{H}^1 \subseteq \mathbb{R}^{N \times d_1}$, where $d_1$ is the first embedding dimension:

$$\mathbf{H}^1 = \sigma\left(\widehat{\mathbf{A}} \mathbf{X} \mathbf{W}^{(1)}\right) = [\mathbf{h}_i^1]_{i=1}^N \subseteq \mathbb{R}^{d_1} \tag{9}$$

where $\mathbf{W} \in \mathbb{R}^{F \times F'}$ is the weight matrix for the convolution operation. This process is repeated for the subsequent layers:

$$\mathbf{H}^{l+1} = \sigma\left(\widehat{\mathbf{A}} \mathbf{H}^l \mathbf{W}^{(l+1)}\right) = [\mathbf{h}_i^{l+1}]_{i=1}^N \subseteq \mathbb{R}^{d_{l+1}} \tag{10}$$

After $L$ layers, the output is $\mathbf{H}^L \subseteq \mathbb{R}^{N \times d_L}$, where $d_L$ is the final embedding dimension. To predict the output $y$ directly from the final node embedding $\mathbf{H}^L$, we apply a fully connected layer to the aggregated node embeddings. The prediction $\hat{y}$ can be represented as:

$$\hat{y} = \sigma(\mathbf{W}_{\text{out}} \mathbf{H}^L + \mathbf{b}_{\text{out}}) \tag{11}$$

where $\mathbf{W}_{\text{out}} \in \mathbb{R}^{d_L \times 1}$ is the weight matrix for the output layer; $\mathbf{b}_{\text{out}} \in \mathbb{R}^1$ is the bias term.

The total loss for the model is defined as:

$$J(\hat{y}_i, y_i) = \frac{1}{N} \sum_{i=1}^{N} (\hat{y}_i - y_i)^2 \quad (12)$$

where $\hat{y}_i$ is the predicted porous medium property from the last layer, and $y_i$ is the ground truth porous medium property. In GNN, the backpropagation algorithm is utilized to update the weights by computing gradients of the loss function $J$ with respect to the parameters $\theta$. The weight update rule for gradient descent is given by:

$$\theta^{(t+1)} = \theta^{(t)} - \eta \frac{\partial J}{\partial \theta} \quad (13)$$

where $t$ represents the iteration number in the optimization process during gradient descent; $\eta$ is the learning rate. The gradients for the weights are computed using the chain rule:

$$\frac{\partial J}{\partial \mathbf{W}^{(l)}} = \frac{\partial J}{\partial \mathbf{h}_v^{(l+1)}} \frac{\partial \mathbf{h}_v^{(l+1)}}{\partial \mathbf{W}^{(l)}} \quad (14)$$

Thus, we have:

$$\frac{\partial J}{\partial \theta} = \sum_l \frac{\partial J}{\partial \mathbf{W}^{(l)}} \quad (15)$$

where $\mathbf{W}^{(l)}$ are the weights at the layer $l$.

Following the development of the GNN and 2D CNN models, we introduced a fusion model, named the GNN-2D CNN, which seamlessly integrates the capabilities of both architectures (**Fig. 4**). Fusion models, which merge multiple input sources or different feature representations, have made significant impacts in computer vision applications, especially in scenarios involving multimodal images or different image sensors (Jones et al., 2021). Such fusion techniques exploit the complementary nature of multiple feature representations, thereby providing a more comprehensive insight into the data. Furthermore, by merging different representations or modalities, fusion models increase robustness and reduce the uncertainties inherent in each individual representation. Based on these advantages, we propose a hybrid neural network that merges feature representations from the separately trained GNN and 2D CNN models. While the GNN is designed to handle node and edge features of graphs, the 2D CNN specializes in processing pixel-level information. After processing their respective inputs, the resulting outputs from both models are concatenated and then passed through dense layers for the final porous medium property predictions.

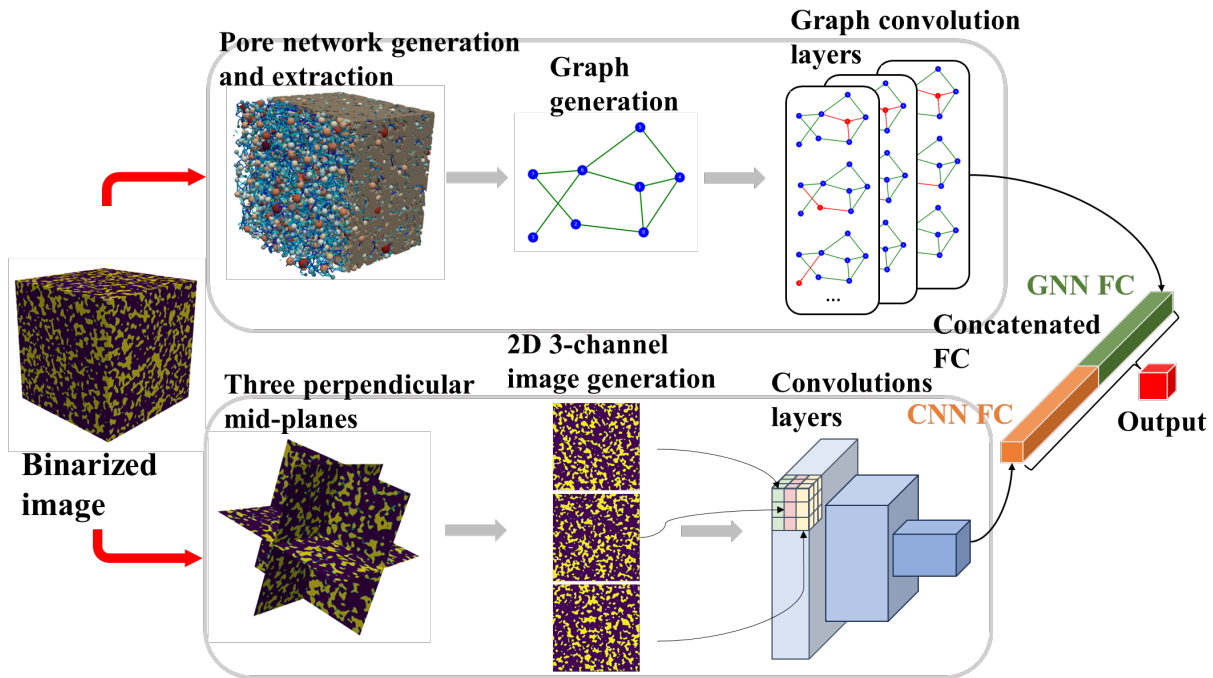

**Fig. 4**. Schematic diagram of the fusion model integrating GNN and 2D CNN for porous media property prediction. The upper panel shows the GNN pathway, where a 3D pore structure is transformed into a graph that represents pores (nodes) and throats (edges). The lower panel illustrates the 2D CNN pathway, where the 3D image is sliced into 2D cross-sections to capture local details. The outputs of the GNN (green bar) and CNN (orange bar) are concatenated through a fully connected layer (FC), enhancing prediction accuracy by leveraging both structural and spatial information.

## 2.4. Gradient-Weighted Class Activation Mapping

The gradient-weighted class activation mapping (Grad-CAM) technique (Selvaraju et al., 2017) is employed to identify the influential regions in porous media images for the significant nodes that affect porous medium property predictions in the GNN model. Grad-CAM is a technology designed to provide visual explanations for decisions made by deep learning models, particularly for CNNs. For the GNN model, an adapted version of Grad-CAM was employed, focusing on the significance of individual nodes within a graph (**Fig. 5**). It resulted in the creation of a "node map", highlighting the pivotal nodes influencing the porous medium property predictions. First, we performed a forward pass through the model while keeping track of the node embeddings and enabling gradient computation for the input features. Then, a backward pass was conducted from a target node. The gradient of the node embeddings was computed and utilized to derive importance weights for each node. These importance weights were then visualized to identify which parts of the graph (or which nodes) were crucial for the model's predictions. For visualization purposes, the pore network was plotted and overlaid onto the original image. The GNN Grad-CAM scores were set as different colors for each pore, providing a clear visual indication of the nodes that significantly influenced the porous medium property predictions.

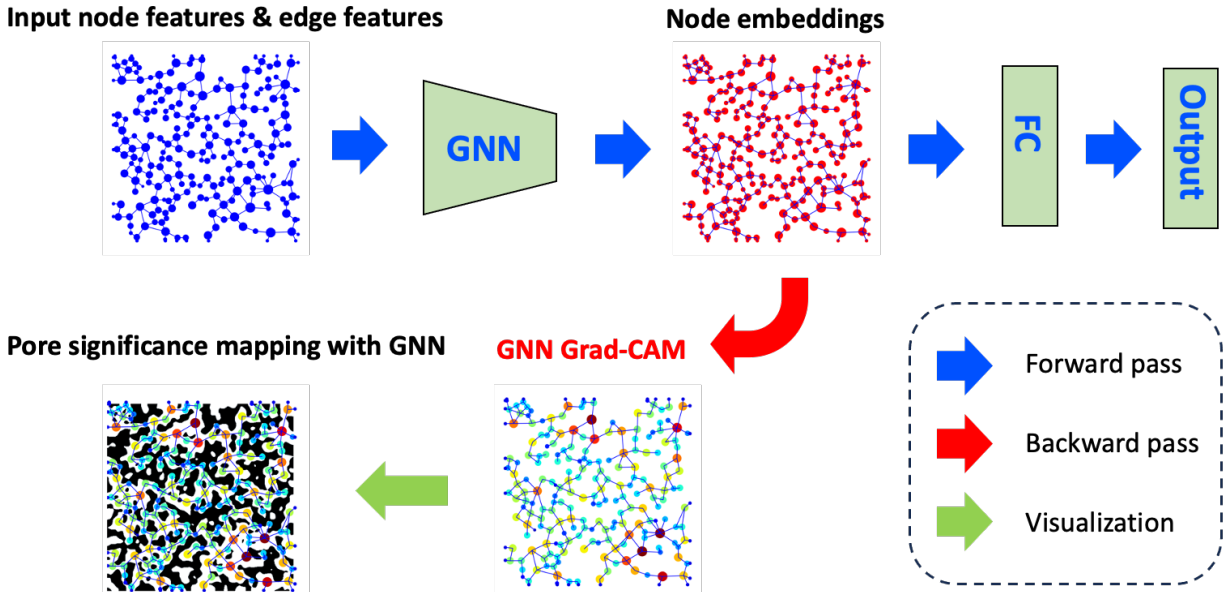

**Fig 5**. Workflow of the Grad-CAM adapted for GNNs, emphasizing the nodes in the graph crucial for the model's predictions. FC indicates fully connected layers.

## 3. Results and Discussion

### 3.1. Comparison of the Model Performance

To evaluate the performance of different deep learning models — the 2D CNN, 3D CNN, GNN, and the GNN-2D CNN fusion model, we utilized two porous medium properties: permeability and formation factor, from a testing dataset comprised of 1,770 rock images (Rabbani et al., 2020). We utilized $R^2$ score, mean absolute error (MAE), and root mean square error (RMSE) as the model metrics to quantify the model accuracy. **Fig. 6** and **7** show the porous medium properties predictions. For permeability predictions, the GNN demonstrated superior performance compared to the 2D CNN. The 2D CNN had the lowest prediction accuracy among the four models presented. This is because only limited compressed information is extracted from the original 3D images, which results in a considerable loss of information. The comparison between the 2D CNN and GNN highlights that the use of a graph data structure for image simplification is an effective approach, as the GNN captures relational features between pores and throats that are directly relevant to permeability.

Given that both models use simplified representations of porous media images as inputs, the GNN and 2D CNN models exhibited lower accuracy than the 3D CNN, which utilized the complete images as inputs. The 3D CNN demonstrated robust performance for porous medium permeability predictions, indicating that the information from the full-size image is crucial for deep learning prediction accuracy. However, the GNN showed better prediction performance for larger permeability values. This discrepancy likely arises because CNNs are more sensitive to the skewness of the data distribution, where the long tail toward higher values poses a challenge. The CNN models tend to optimize accuracy for the more densely populated middle-range values, sacrificing the prediction performance for the extreme values.

The GNN-2D CNN fusion model demonstrated prediction accuracy comparable to the 3D CNN, highlighting the complementarity of these two approaches. By integrating detailed 2D image features from CNNs with the relational and topological insights provided by GNNs, the hybrid model captures both fine-scale spatial details and large-scale network structures. This combination enables efficient and accurate predictions by leveraging CNNs' ability to extract detailed, pixel-level features and GNNs' strength in representing complex, large-scale relationships, such as the connectivity between pores and throats that is highly correlated with permeability. The hybrid approach illustrates that multimodal, reduced-dimensional models can achieve performance comparable to full 3D models while being more computationally efficient, making it a promising tool for predicting properties in complex porous media.

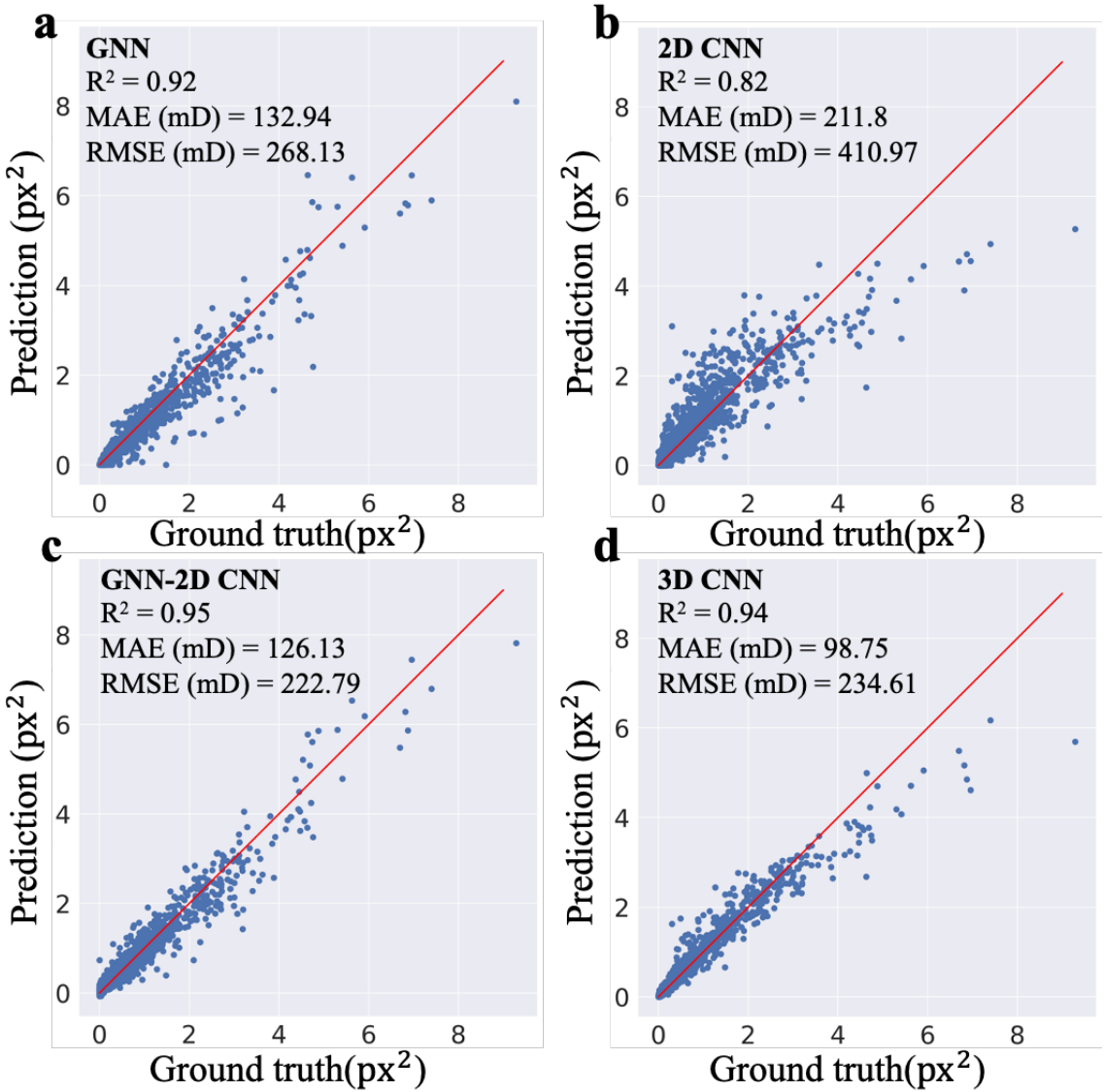

**Fig. 6**. Performance and efficiency comparisons between (a) GNN, (b) 2D CNN, (c) GNN-2D CNN, and (d) 3D CNN in permeability predictions. The MAE and RMSE are calculated by converting the predicted permeability from pixel² to millidarcy (mD), assuming a resolution of 1 μm/pixel.

**Table 5** presents a comparative overview of existing neural network models used for predicting permeability of digital rock images. Although these models were trained and validated on different datasets, which makes direct comparisons challenging, the table highlights general trends in performance improvements over time. Notably, many recent models have achieved high $R^2$ scores, reflecting advances in neural network architectures and training methodologies. Our proposed hybrid model, which integrates GNNs with 2D CNNs, achieves a $R^2$ score of 0.95. This performance is comparable to state-of-the-art 3D CNN models while benefiting from reduced memory requirements. By leveraging the complementary strengths of GNNs for capturing relational features and 2D CNNs for efficient image processing, our approach demonstrates that hybrid models can offer a balance between accuracy and computational efficiency. This makes our model particularly attractive for applications where memory resources are limited.

**Table 5.** Overview of neural network models for image-based permeability predictions.

|  | Method | $R^2$ | MAE (mD) | RMSE (mD) |
|---|---|---|---|---|
| Current model | GNN, 2D CNN | 0.95 | 126.13 | 222.79 |
| J. Wu et al. (2018) | 2D CNN | 0.92 | N/A | N/A |
| Hong & Liu (2020) | 3D CNN | 0.92 | N/A | 761.1174 |
| Tembely et al. (2020) | 3D CNN | 0.91 | N/A | N/A |
| Kamrava, Sahimi, et al. (2021) | 3D CNN | 0.91 | N/A | N/A |
| Alqahtani et al. (2021) | 3D CNN | 0.87 | 40 | N/A |
| Tang et al. (2022) | 3D CNN | 0.99 | N/A | N/A |
| H. Zhang et al. (2022) | 2D CNN | 0.9 | N/A | N/A |
| Elmorsy et al. (2023) | 3D CNN | 0.95 | N/A | N/A |
| Alzahrani et al. (2023) | GNN, 3D CNN | 0.91 | 45.1 | N/A |
| Liu et al. (2023) | 3D CNN | 0.99 | N/A | N/A |

| Zhai et al. (2024b) | 2D CNN | 0.97 | N/A | N/A |

Fig. 7 illustrates the models' predictions on formation factors. As with the prediction of permeability, the performance of the model exhibits a comparable trend. The hybrid model exhibits the highest degree of accuracy among the four proposed models. Nevertheless, the prediction accuracy for the formation factor is inferior to that for permeability when the same model structures are employed, consistent with the findings of the DeePore framework (Rabbani et al., 2020). This reduced accuracy can be attributed to the complex and less direct relationship between the formation factor and the pore structure features captured in the images. Unlike permeability, which is strongly correlated with pore and throat diameters, the formation factor depends more intricately on the connectivity and tortuosity of the pore network, making it harder to infer from image-based features alone. The inherent variability and sensitivity of the formation factor to small-scale structural heterogeneities further complicate accurate predictions.

In contrast to the performance observed in permeability predictions, the hybrid model demonstrated superior performance in formation factor predictions, outperforming the 3D CNN model. The improved performance of the hybrid model suggests that combining multiscale features from both graph-based and image-based representations can effectively capture more complex structural attributes influencing the formation factor. However, the single GNN and 2D CNN models exhibited considerably lower prediction accuracy than the 3D CNN model. This indicates that although features extracted from a simplified model cannot match the performance of the full-information 3D CNN model, multiscale feature fusion technology can significantly enhance the simplified model's capability. Overall, the prediction results emphasize the need for models that can integrate detailed connectivity and tortuosity features to further improve formation factor predictions.

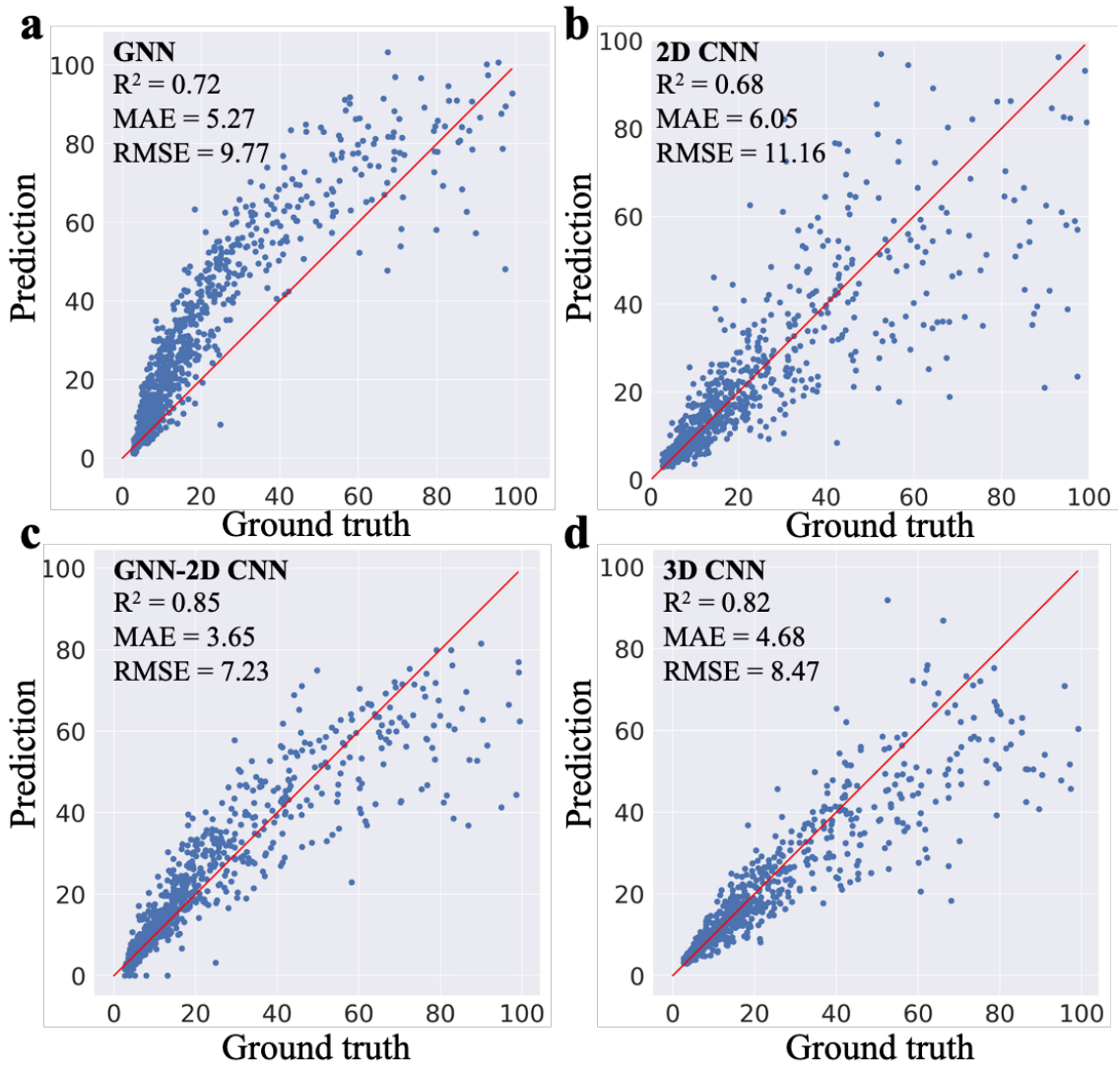

**Fig 7**. Performance and efficiency comparisons between the (a) GNN, (b) 2D CNN, (c) GNN-2D CNN, and (d) 3D CNN in formation factor predictions.

**3.2. Interpretability of the Hybrid Model for Porous Medium Property Predictions**

GNNs offer a high degree of interpretability, which is advantageous for predicting the properties of porous media. The pore network inherently resembles a graph structure, with nodes representing individual pores and edges representing the connections between them. Each node is characterized by features such as pore diameter and pore volume. To illustrate the interpretability of GNNs for porous medium property predictions, we utilized Grad-CAM. This technique enabled us to effectively visualize the contribution of various pores to the prediction outcomes.

**Fig. 8a** shows the GNN Grad-CAM results for the GNN permeability prediction with graphs extracted from 3D image data. As shown in the figure, the Grad-CAM assigns different scores to different pores. These scores represent the gradients of node features concerning the prediction results, indicating the importance of each node in the predictions. Node features, such as pore diameter, pore volume, and pore surface area, are directly used in the gradient calculation. Furthermore, the features of connected edges of the nodes are also considered in the node features by our GNN model structure, by concatenating the edge features and node features before graph convolution. Edge features, including edge diameter, edge length, and other important parameters, are included in the model training process. Another crucial factor of porous media is connectivity, which denotes the number of adjacent pores to a given pore. Higher average pore connectivity generally leads to higher permeability (Alhashmi et al., 2016; Blunt, 2001; Rabbani et al., 2020). In this study, the connectivity between these pores is accounted for implicitly within the node features. By using sum aggregation during the convolutional process, as shown in Equation 7, the intrinsic connectivity information within the graph structure was preserved.

In contrast to traditional CNNs, GNNs offer a more interpretable framework for this specific application. As CNN layers deepen, interpreting the resulting feature maps becomes increasingly challenging due to the abstracted and high-dimensional nature of the learned features. However, GNNs maintain a fixed graph size during the aggregation process, facilitating consistent visualization of gradients across different parts of the porous media.

Our approach did not involve any graph pooling, thereby preserving the original graph structure throughout the network. This decision was crucial for maintaining the interpretability of the model, as graph pooling could obscure the direct relationships between individual pores and their contribution to the overall prediction. By applying Grad-CAM, we generated heatmaps that highlight the regions within the pore network that are most influential in predicting permeability. These heatmaps provided clear visual insights into which pores play a significant role in the model's predictions. The fixed graph size ensured that these visualizations remained consistent and readily interpretable, regardless of the depth of the GNN layers. The ability to directly observe and analyze the importance of different pores within the network underscores the advantage of using GNNs for this task. This method not only improves prediction accuracy but also facilitates a deeper comprehension of the underlying factors influencing permeability of the porous media.

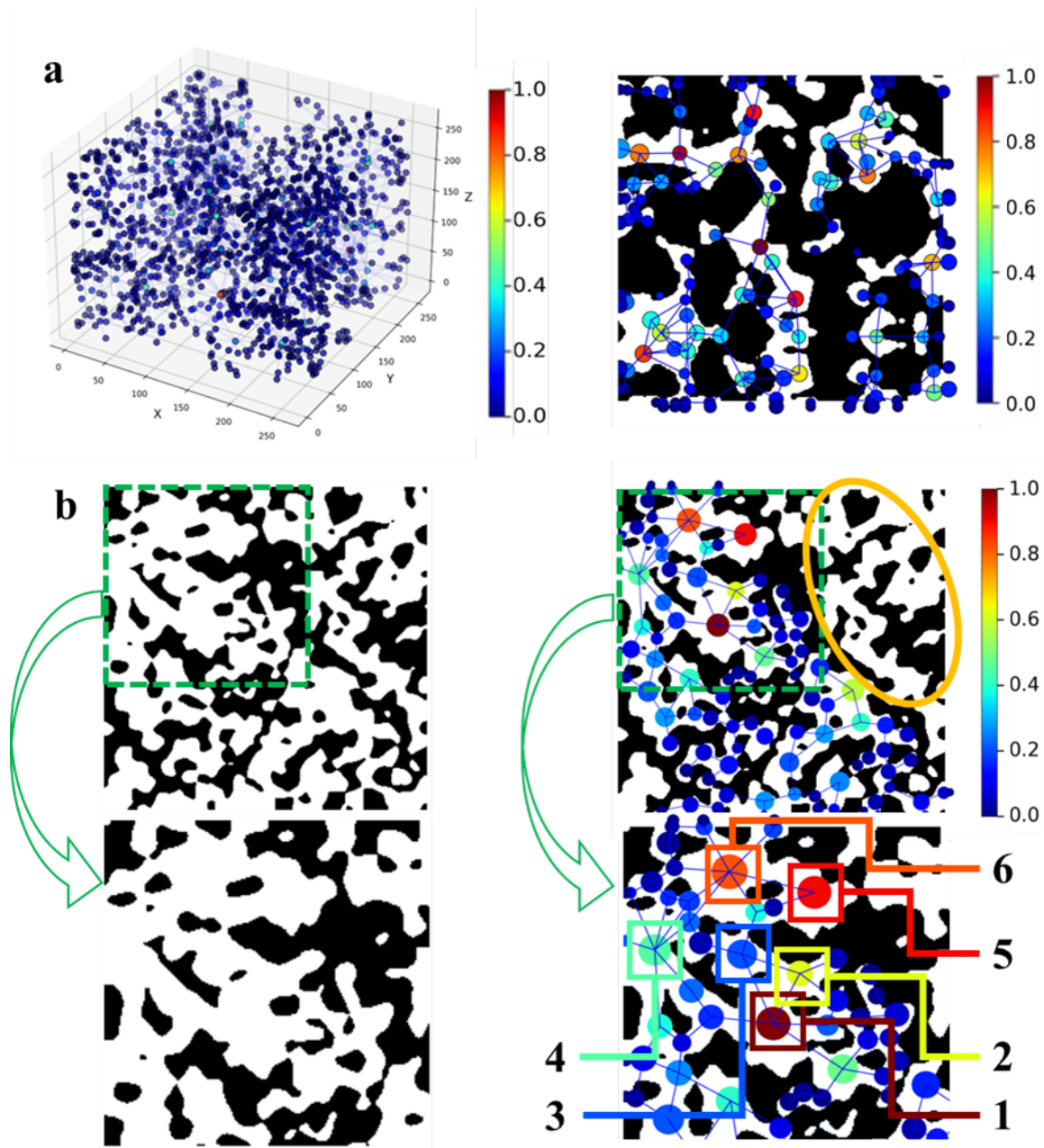

**Fig. 8**. Grad-CAM-based analysis of the GNN's accuracy in predicting permeabilities across diverse pore geometries from (a) 3D image, (b) 2D image.

Although the 3D graph in Fig. 8a provides a clear relationship between each pore and the permeability prediction, interpreting the model in a 2D paper remains challenging. This difficulty arises because a 2D slice of the 3D graph cannot be perfectly aligned with the porous medium image. As illustrated in the right figure of Fig. 8a, some nodes may exhibit elevated Grad-CAM scores despite exhibiting reduced connectivity and pore volume. This observation can be attributed to the fact that some pores oriented in a direction parallel to the plane of the paper are not visible. The existence of these "invisible" pores in three-dimensional space has the potential to disrupt the interpretation of the model. To circumvent this issue, we generated 2D Grad-CAM results for the GNN model, as illustrated in **Fig 8b**.

In practice, "died pore" is a common phenomenon in porous media, where fluid flow is impeded due to the absence of a direct connection to the main body of the porous media. It is possible that neural networks can not distinguish the "died pores" explicitly. To address this issue, the disconnected pores were removed to ensure a fully connected pore network during the PNM extraction process. The GNN's Grad-CAM result in Fig. 8b displays no nodes in the top-right corner, where these pores are isolated from the main body of the 2D porous medium. This method ensures that the GNN can accurately identify and consider only the effective, connected pores, thereby enhancing the prediction precision.

We then investigated which specific regions the GNN model focused on while making predictions. For a more detailed analysis, we focused on the top-left section of Fig. 8b. The zoom-in GNN's Grad-CAM provides a clear relationship between the pore connectivity and the intensity of the importance. In this detailed representation, six nodes were specifically selected for the examination. Upon closer inspection, Node 1 stands out with the highest intensity. It is surrounded by five neighboring nodes, of which three have an intensity of no less than 0.4. The node intensity accounts for the connectivity of the pore, its inherent properties such as the pore size, and the unique properties of its adjacent pores. Nodes 5 and 6, despite being similar to Node 1 in size, have less intensity than it. This is attributed to the fact that Node 5 is connected to only two neighbors, and the neighbors of Node 6 have a less significant role than those of Node 1. In addition, Nodes 2, 3, and 4 have intermediate intensities, which is due to their smaller dimensions and other intrinsic properties.

Even though the six pores circled by boxes in Fig. 8b have similar pore volume and connectivity, their colors are different, implying that the GNN model assigns different levels of importance to them for the prediction. This certifies that GNN is able to evaluate the contributions of different pores on the prediction, depending on the pore's connectivity and the properties of the connected pores. This process enables GNN to comprehend the relationships between each pore and its neighbors, thereby facilitating the understanding of flow pathways, connectivity, and pore interactions.

### 3.3. Efficiency of GNNs for Porous Medium Property Predictions

By upscaling information from the pixel level to the pore network level, the graph structure provides a more efficient way to represent porous media. The training time per epoch and the total number of trainable parameters for each proposed model were calculated and are presented in **Table 6**. The GNN has the fewest total trainable parameters, but a higher training time per epoch compared to the 2D CNN. This difference may be attributed to the data structure and different model mechanisms. The 3D CNN has the greatest training time per epoch and the highest total number of trainable parameters due to the extensive input data dimension and the 3D convolution layers necessitated by 3D data. **Fig. 9** illustrates the accuracy and total number of parameters for the proposed models. Additionally, the utilization of GPU memory is presented and denoted by the size of the circles. It is evident that the GNN provides relatively high accuracy among all models, with only negligible GPU memory usage and a small number of total parameters. The GNN-2D CNN fusion model, which integrates the 2D CNN and GNN input features, exhibits performance comparable to that of the 3D CNN in permeability predictions, yet with considerably lower GPU memory usage and total parameter count. The GPU memory usage and total number of parameters of the fusion model are nearly identical to those of the 2D CNN model, yet its accuracy is demonstrably superior. This suggests that the GNN is a vital component of the hybrid model for porous medium permeability predictions. This plot highlights the outstanding efficiency of the GNN model for predicting porous medium properties. This efficiency is primarily due to the natural representational capabilities of graphs for porous media and the methodology employed for upscaling from high-resolution, memory-intensive images to low-resolution, memory-efficient graphs. The efficiency of the graph representation and the GNN model demonstrates promise for the larger-scale porous medium image prediction and generation tasks, which can be particularly time-consuming with existing numerical technologies.

**Table 6.** Computational accuracy and efficiency of neural network models on porous medium permeability predictions.

| Model | $R^2$ | MAE (mD) | RMSE (mD) | Training time/epoch (s) | Total trainable parameter number |
|---|---|---|---|---|---|
| GNN | 0.92 | 132.94 | 268.13 | 7.9 | 9,248 |
| 2DCNN | 0.82 | 211.8 | 410.97 | 3.8 | 798,325 |
| GNN-2DCNN | 0.95 | 126.13 | 222.79 | 16.9 | 812,692 |
| 3DCNN | 0.94 | 98.75 | 234.61 | 574.7 | 50,342,629 |

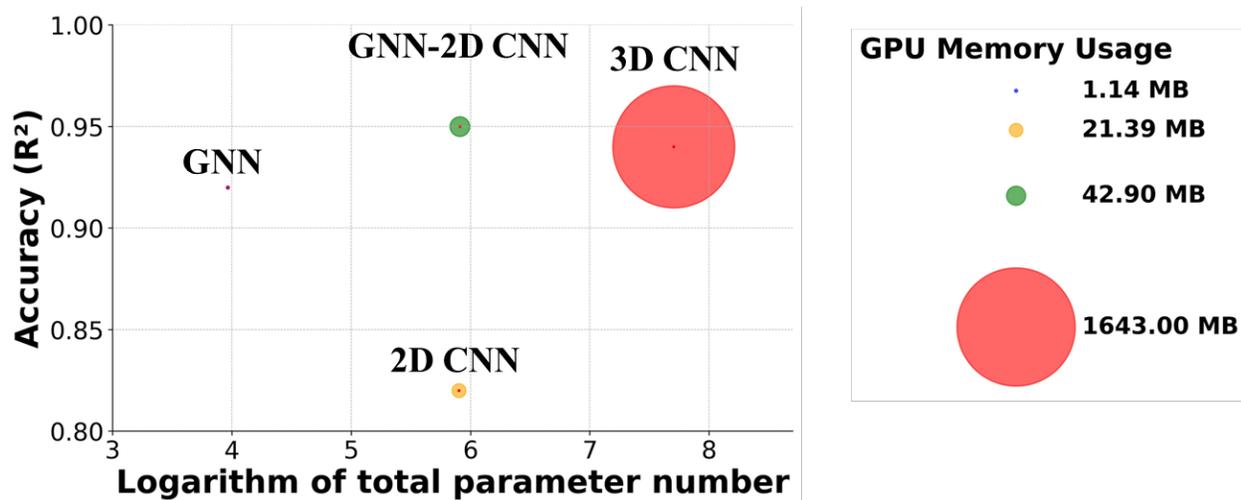

**Fig. 9**. Performance and efficiency comparisons between the GNN, 2D CNN, GNN-2D CNN, and 3D CNN in permeability predictions. The model accuracy is plotted against the logarithm of the total parameter number; the circle size indicates GPU memory usage.

To further illustrate the efficiency of the GNN model, we analyzed how trainable parameters scale with image size across different models, as illustrated in **Fig. 10**. For GNNs, the number of trainable parameters remains constant regardless of image or node size, as it depends solely on the number of node and edge features, not on the image dimensions. In the GNN model, each node and edge is associated with a constant number of features that are processed by neural networks. The nodes and edges that are situated within the same convolution layer share the same trainable parameters. This design allows the GNN to maintain a fixed parameter count even as the image size grows. In contrast, the parameters for CNN models, both 2D and 3D, increase significantly as image dimensions grow, resulting in high memory demands for large images. The GNN-2D CNN hybrid model has a nearly identical parameter count to the 2D CNN alone because the additional parameter number from the GNN component is negligible.

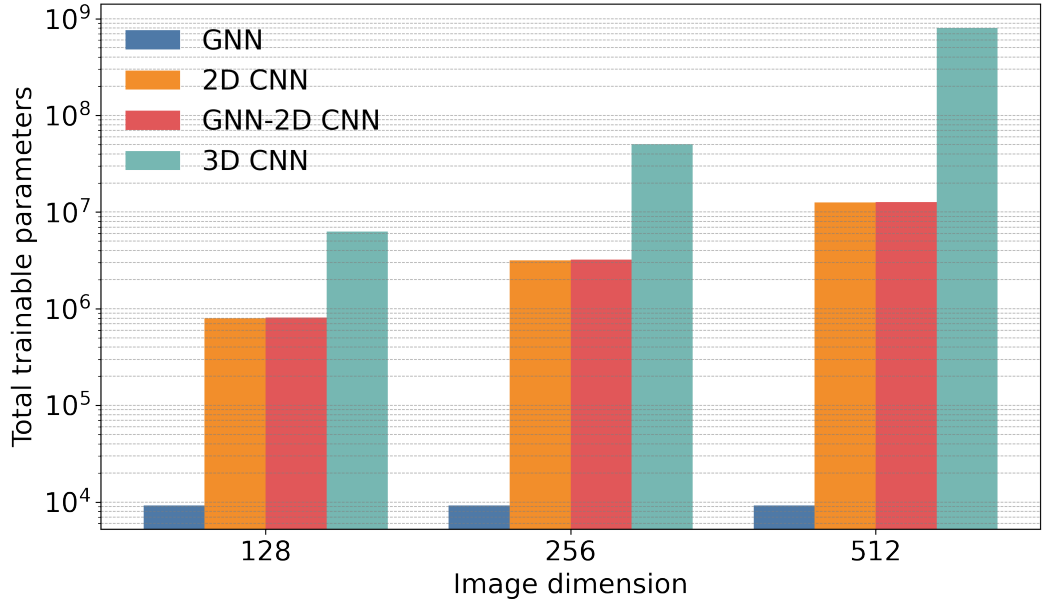

**Fig. 10.** Number of trainable parameters of different models (2D CNN, GNN, GNN-2D CNN, and 3D CNN) against the image dimension. The 2D CNN processes images with dimensions of 128×128×3, 256×256×3, and 512×512×3, whereas the 3D CNN processes images with dimensions of 128×128×128, 256×256×256, and 512×512×512.

## 4. Conclusions

We present a novel GNN approach that demonstrates significantly greater memory efficiency compared to traditional CNN models. This efficiency is achieved by consolidating clusters of pixels into nodes and edges representing pores and throats, respectively. The graph data structure is naturally aligned with the porous medium structure, rendering GNN a promising tool for the prediction of porous medium properties. By employing the innovative GNN Grad-CAM technology, we have provided insights into the mechanisms of GNN and their effectiveness in understanding fluid dynamics in porous media, offering interpretability that was previously lacking. Our work highlights the potential of GNN models for scalable simulations of fluid flow in porous media, addressing challenges that existing methods struggle with. Furthermore, the GNN model can be employed to generate large-dimensional porous media materials for applications in medicine and material design, with relatively lower memory demands. A key contribution of this study is the introduction of a fusion model, an innovative methodology that adeptly leverages the strengths of both CNN and GNN for multiscale feature extractions. This integration seamlessly combines the strengths of both the CNN and the GNN in handling intricate spatial details and relational data, respectively. The hybrid model's efficient design enables high prediction accuracy with reduced computational resources, representing a pioneering effort in the field of property predictions in porous media. The efficacy of this model underscores the potential of hybrid neural network architectures to exceed current limits in digital rock physics and related disciplines.

**Acknowledgments**

The authors are thankful for the financial support provided by the U.S. Department of Energy (DOE)'s Nuclear Energy University Program (NEUP) through the Award Number of DE-NE0009160. The authors express their gratitude to the creators of DeePore for providing the porous material dataset, which was invaluable to this study.

**Data Availability Statement**

The porous media image datasets used in this study are available through: https://doi.org/10.1016/j.advwatres.2020.103787, and https://doi.org/10.5281/zenodo.3820900. The graph datasets are available through: https://doi.org/10.5281/zenodo.12786239.